# Self-assembled cyclic oligothiophene nanotubes: electronic properties from a dispersion-corrected hybrid functional

Bryan M. Wong[1],* and Simon H. Ye[2]

[1]*Materials Chemistry Department, Sandia National Laboratories, Livermore, California 94551, USA*

[2]*Department of Chemistry, Stanford University, Stanford, California 94309, USA*

*Corresponding author; bmwong@sandia.gov

The band structure and size-scaling of electronic properties in self-assembled cyclic oligothiophene nanotubes are investigated using density functional theory (DFT) for the first time. In these unique tubular aggregates, the $\pi$-$\pi$ stacking interactions between adjacent monomers provide pathways for charge transport and energy migration along the periodic one-dimensional nanostructure. In order to simultaneously describe both the $\pi$-$\pi$ stacking interactions and the global electronic band structure of these nanotubes, we utilize a dispersion-corrected B3LYP-D hybrid functional in conjunction with all-electron basis sets and one-dimensional periodic boundary conditions. Based on our B3LYP-D calculations, we present simple analytical formulas for estimating the fundamental band gaps of these unique nanotubes as a function of size and diameter. Our results on these molecular nanostructures indicate that all of the oligothiophene nanotubes are direct-gap semiconductors with band gaps ranging from 0.9 eV – 3.3 eV, depending on tube diameter and oligothiophene orientation. These nanotubes have cohesive energies of up to 2.43 eV per monomer, indicating future potential use in organic electronic devices due to their tunable electronic band structure and high structural stability.

PACS number(s): 61.46.Np, 71.15.-m, 73.22.-f, 78.67.Ch

## I. INTRODUCTION

Nanostructures consisting of conjugated thiophene chains are one of the most frequently studied classes of photovoltaic nanomaterials due to their highly conjugated $\pi$-bonding systems, chemical stability, and tunable electronic properties.[1] Because of their high carrier mobilities, oligo- and polythiophenes have been utilized in organic field-effect transistors (OFETs), organic light-emitting diodes (OLEDs), and photovoltaic materials.[2,3] Although linear thiophene chains have shown great utility in optoelectronic devices, cyclic organic materials are of considerable interest due to their conserved and symmetrical three-dimensional structures. For example, porphyrin nanotubes are known to self-assemble via strong ionic interactions,[4] and cyclic oligothiophenes have also demonstrated very stable self-assembling properties.[5] Furthermore, recent synthetic advances[6] in the creation of various cyclic oligothiophenes[7-9] have opened up the possibility of forming oligothiophene-based nanostructures with specific sizes and chemical functional groups.[10] Compared to conventional carbon nanotubes bonded via strong covalent interactions, cyclic oligothiophene nanotubes are held together along the tube axis via purely noncovalent interactions, lending to facile self-assembly from individual cyclic monomers.[11] In addition, since the electronic interactions between adjacent nanotubes can be selectively tuned via chemical functionalization with side chains,[12] these materials, in principle, can be tailored to modulate quasi one-dimensional electronic transport along the tube axis.[13-15] As a result, self-assembled cyclic oligothiophene nanotubes are potentially a new class of organic nanotubes with tunable electronic properties which can be utilized as semi-conducting materials in nano-electronic devices.

In order to quantitatively predict the electronic band structure of these cyclic oligothiophene nanotubes, which have not been previously investigated in the framework of fully-periodic boundary conditions, the use of density functional theory (DFT) as a first-principles tool is a natural choice. However, in choosing a specific method within the DFT formalism, one must be cautious by recognizing two well-known shortcomings of conventional functionals, especially in the context for calculating the unique electronic structure of noncovalently-bound nanostructures. First and most importantly, DFT methods utilizing the local density approximation (LDA) or the generalized gradient approximation (GGA) systematically underestimate band gaps in semiconductors, insulators, and strongly-correlated systems.[16-19] This deficiency arises from spurious electron self-interaction in semi-local functionals and the lack of a derivative discontinuity of the exchange-correlation potential with respect to electron occupancy.[20,21] Hybrid functionals such as B3LYP, which incorporate a portion of nonlocal Hartree-Fock (HF) exchange, partially ameliorate the self-interaction problem and produce more accurate band gaps than the LDA or GGA approaches.[22-24] A striking example of this improved accuracy can be found in the recent study by the Goddard group which showed that both LDA and GGA approaches predict very small band gaps in single-wall carbon nanotubes.[25] Furthermore, these researchers found that the B3LYP hybrid functional with periodic boundary conditions (PBC) leads to very accurate band gaps in excellent agreement with experiment.

Another shortcoming of conventional DFT functionals is the poor description of dispersion interactions which, in the case of our noncovalently-bonded nanotubes, are crucial to their stability. Although the B3LYP hybrid functional produces accurate band gaps, it still fails completely for noncovalent interactions.[26-39] The reason for this failure is that many DFT approximations (including B3LYP) do not accurately account for correlation effects describing the instantaneous multipole/induced multipole charge fluctuations between molecular surfaces. As a result, standard B3LYP-geometry optimizations on noncovalently-interacting systems can typically lead to unbound clusters and dissociation of adsorbed species.[29,34-39] One efficient method to include dispersion effects is the DFT-D approach by Grimme[26] which simply adds an empirical, interatomic dispersion-energy contribution to DFT total energies. The main appeal of the DFT-D method is that it can be easily coupled to existing exchange-correlation functionals with a proper re-parameterization of dispersion coefficients. Although the DFT-D formalism requires two empirical parameters for every element, this approach has given very accurate results for numerous intermolecular interactions benchmarked by high-level wavefunction based approaches (i.e., Møller-Plesset second-order perturbation theory or the coupled cluster method).[40-43] It is also important to mention at this point that there are other less empirical approaches for including dispersion effects which have attracted considerable attention in the last few years. *Ab initio* methods such as adiabatic-connection fluctuation-dissipation (ACFD) approaches[44,45] and exact exchange with a random-phase approximation for the correlation energy (EX+cRPA)[46-48] are still very computationally intensive and can only be applied to small systems. Alternatively, one of the more well-known approaches is the nonlocal van-der-Waals density functional (vdW-DF) due to Langreth and Lundqvist.[49-51] Although original calculations with vdW-DF were computationally intensive, recent implementations of this nonlocal functional no longer scale unfavorably with system size, making vdW-DF calculations now feasible for systems greater than 100 atoms.[52] There has also been recent work in modifying the vdW-DFT approach for noncovalent interactions in molecular systems due to Vydrov and Van Voorhis.[53,54] Finally, Tkatchenko and co-workers have presented a new scheme to obtain accurate van der Waals interactions from DFT and empirical-free atom reference data[55] which has also been combined with hybrid functionals.[56]

In this work, we investigate the band structure and size-scaling of electronic properties in self-assembled cyclic oligothiophene nanotubes using a dispersion-corrected B3LYP-D hybrid functional (benchmark comparisons of B3LYP-D against *ab initio* vdW-DF calculations[57] are first presented in Section III to validate our chosen approach). It is important to mention that there has also been recent work in carbozole macrocycles[12] and a similar study on cyclic oligothiophene multimers[58] using different theoretical methods. However, both of these studies focused only on isolated molecular aggregates and did not address band-structure properties in a fully-periodic nanotube geometry. As a result, their calculations are only appropriate for molecular systems and do not capture the full electronic band structure as a function of electron momentum (i.e., molecular calculations are incapable of determining whether a material has a direct (or indirect) band gap, which is an essential property for describing optoelectronic and electron-transport efficiencies in these nanotubes). Indeed, the use of fully-periodic approaches for an accurate description of electronic features (band structure and gap) is mandatory since the modeling of extended systems using clusters can introduce spurious border effects related to the finite size of the multimers, potentially affecting the representation of the band structure (especially the conduction band[59,60]). It is also important to point out that the previous molecular study by Flores[58] used the MPWB1K hybrid functional to calculate noncovalent binding energies; however, a recent study by Grimme[61] has shown that the MPWB1K functional (as well as newer versions of MPWB1K such as M05-2X and M06-2X) still does not recover the correct long-range $R^{-6}$ dispersion energy as a function of internuclear distance. Since functionals such as MPWB1K neglect the long-range dispersion energy, Grimme and co-workers have found that they yield significantly smaller binding energies in large carbon systems, and that dispersion-corrected functionals such as B3LYP-D[26] are essential for describing $\pi$-$\pi$ stacking interactions in these systems. As a result, we have chosen the B3LYP-D functional, in conjunction with all-electron basis sets and one-dimensional periodic boundary conditions, to carry out an accurate description of both the $\pi$-$\pi$ stacking interactions and the global electronic band structure in our nanotubes. Following benchmark calculations on polythiophene and comparisons with other *ab initio* studies in the literature, we then examine the effect of nanotube diameter and oligothiophene orientation on their stability and electronic properties. We begin by briefly describing the B3LYP-D approach and then discuss its implications for tuning the electronic and geometric properties of these tubular nanostructures.

## II. THEORY AND METHODOLOGY

Within the DFT-D approach, an empirical atomic pairwise dispersion correction is added to the Kohn-Sham part of the total energy ($E_{\text{KS-DFT}}$) as

$$E_{\text{DFT-D}} = E_{\text{KS-DFT}} + E_{\text{disp}}, \qquad (1)$$

where $E_{\text{disp}}$ is given by

$$E_{\text{disp}} = -s_6 \sum_{i=1}^{N_{\text{at}}-1} \sum_{j=i+1}^{N_{\text{at}}} \sum_{\mathbf{g}} f_{\text{damp}}\left(R_{ij,\mathbf{g}}\right) \frac{C_6^{ij}}{R_{ij,\mathbf{g}}^6}. \qquad (2)$$

Here, the summation is over all atom pairs $i$ and $j$, and over all $\mathbf{g}$ lattice vectors with the exclusion of the $i = j$ contribution when $\mathbf{g} = 0$ (this restriction prevents atomic self-interaction in the reference cell). The parameter $C_6^{ij}$ is the dispersion coefficient for atom pairs $i$ and $j$, calculated as the geometric mean of the atomic dispersion coefficients:

$$C_6^{ij} = \sqrt{C_6^i C_6^j}. \qquad (3)$$

The $s_6$ parameter is a global scaling factor which is specific to the adopted DFT method ($s_6$ = 1.05 for B3LYP), and $R_{ij,\mathbf{g}}$ is the interatomic distance between atom $i$ in the reference cell and $j$ in the neighboring cell at distance $|\mathbf{g}|$. A cutoff distance of 25.0 Å was used to truncate the lattice summation which corresponds to an estimated error of less than 0.02 kJ/mol on cohesive energies, as determined by previous studies.[36] In order to avoid near-singularities for small interatomic distances, the damping function used in Eq. (1) has the form

$$f_{\text{damp}}(R_{ij,\mathbf{g}}) = \frac{1}{1+\exp\left[-d\left(R_{ij,\mathbf{g}}/R_{\text{vdW}}-1\right)\right]}, \quad (4)$$

where $R_{\text{vdW}}$ is the sum of atomic van der Waals radii $\left(R_{\text{vdW}} = R_{\text{vdW}}^i + R_{\text{vdW}}^j\right)$, and $d$ controls the steepness of the damping function.

All calculations were carried out with the CRYSTAL09 program,[62] which uses both all-electron Gaussian-type orbitals and exact Hartree-Fock exchange within periodic boundary conditions. Electronic structure calculations for all of the oligothiophene nanotubes utilized the B3LYP-D hybrid functional with dispersion coefficients taken from the original benchmark study by Grimme.[26] We are aware of a very recent re-parameterization of the B3LYP-D coefficients for molecular crystals,[36-39] but we mainly use the original parameters by Grimme since they have been thoroughly benchmarked on several thiophene systems including thiophene-gas complexes ($H_2$, $CO_2$, $CH_4$, and $N_2$)[41,42] and adsorption of thiophene on noble metals (Cu and Au).[43] Geometries for all of the oligothiophene nanotubes were optimized using the 6-31G(d,p) all-electron basis set with one-dimensional periodic boundary conditions along the tube axis. All optimizations were calculated without symmetry constraints, and each unit cell contained two cyclic macrocycles in a parallel-displaced geometry (see Figs. 1 and 4a). At the optimized geometries, a final single-point B3LYP-D calculation was performed with a larger, triple-zeta 6-311G(d,p) basis set to compute the electronic band structure with 100 k-points along the one-dimensional Brillouin zone. Since localized Gaussian basis sets are used in our calculations, the basis set superposition error (BSSE) becomes an issue in evaluating cohesive energies. This particular phenomenon arises from the use of finite-sized basis sets, and in the limit of a complete (infinitely-sized) basis set, the BSSE would be reduced to zero. In our single-point calculations with the large triple-zeta 6-311G(d,p) basis set, we found that the BSSE was negligible when estimated from the counterpoise correction,[63] and that the use of larger or more diffuse basis sets did not significantly improve the electronic wavefunction when periodic boundary conditions were used. As a result, cohesive energies (per monomer) at the B3LYP-D/6-311G(d,p) level of theory were evaluated without the counterpoise correction using the expression

$$E_{\text{cohesive}} = E_{\text{macrocycle}} - E_{\text{tube}}/2, \qquad (5)$$

where $E_{\text{macrocycle}}$ is the total energy of an isolated macrocycle (without periodic boundary conditions), and $E_{\text{tube}}$ is the energy of the unit cell with periodic boundary conditions. The factor of 2 accounts for the number of molecules in the unit cell. According to this definition, the cohesive energy is positive for any stable nanotube.

### III. RESULTS AND DISCUSSION

**Benchmark Calculations.** Since the electronic properties of fully-periodic cyclic oligothiophene nanotubes have not been previously investigated, it is essential to benchmark our methods against high-level cohesive energies and band gaps for known thiophene systems. It should be mentioned that the B3LYP-D method has already shown remarkable accuracy in predicting binding energies in the JSCH-2005 database[64] of 156 noncovalent biological complexes.[65,66] More pertinent to our study is the recent use of the B3LYP-D functional to accurately calculate noncovalent interactions between molecules containing sulfur atoms.[67] To supplement these extensive studies (which focused on only molecular complexes), we also performed additional calculations on bulk organic molecular crystals with noncovalent interactions similar to the nanotubes in our study. Fortunately, the recent publication of full *ab initio* cohesive energies for oligothiophenes by Nabok *et al.*[57] provides an excellent benchmark comparison with our B3LYP-D results. In this previous study, cohesive energies of several oligothiophenes (number of rings, *n* = 2, 4, and 6) in a herringbone packing structure are calculated using the vdW-DF approach. Table I compares our B3LYP-D cohesive energies (computed with the 6-311G(d,p) basis set) against LDA and vdW-DF results, where we take the latter as benchmark reference values. A comparison across each of the oligothiophene monomers indicates that the B3LYP-D results are in excellent agreement with full vdW-DF cohesive energies, with small deviations of only 0.2 eV for the *n* = 4 monomer. The close agreement between the B3LYP-D and vdW-DF results is in stark contrast to the LDA calculations which dramatically underestimate cohesive energies by as much as 30%. The Cartesian coordinates, lattice parameters, and total energies of all our B3LYP-D oligothiophenes in the herringbone packing structure can be found in the Electronic Physics Auxiliary Publication Service (EPAPS)[68] for reference and future studies.

In order to further assess the accuracy of B3LYP-D in predicting solid-state electronic properties, we compute the band gap of periodic polythiophene using LDA, BLYP, and B3LYP-D functionals with the 6-311G(d,p) basis set. As shown in Table II, both the LDA and BLYP functionals severely underestimate the experimentally-determined band gap[69] by nearly 1.0 eV. In contrast, the B3LYP-D band gap is in exceptional agreement with experiment, resulting in a deviation of only 0.05 eV. We should, however, mention that this direct comparison with experiment may be a fortuitous cancellation of several effects. Specifically, the band gap may be different between isolated polymer chains (as calculated here) and for polymer chains in a bulk environment. A proper theoretical treatment would require a GW calculation[70] for the same bulk system (including possible effects such as molecular disorder and defects), which is beyond the scope of the present paper. However, it is clear that the B3LYP-D approach gives more realistic band gaps compared to LDA or GGA, and our benchmark calculations on cohesive energies makes our B3LYP-D approach a reasonable choice for parametric studies on our noncovalently-bound oligothiophene nanotubes.

**Cohesive Energies.** Cyclic oligothiophene nanotubes in both the *syn* and *anti* configurations (Figs. 1a and 1b) were calculated and are denoted as C*n*T-*syn* and C*n*T-*anti*, respectively, where *n* represents the number of thiophene rings in the cyclic monomer. The electronic overlap between monomers in each of these nanostructures is topologically different: the thiophene *p* orbitals are aligned along the axial direction in the *syn* configuration, while the thiopene *p* orbitals point radially outward in the *anti* configuration. In our study, we initially tried several other orientations such as parallel-displaced and perpendicular (T-shaped) geometries, but we found that the *syn* and *anti* configurations gave the most stable one-dimensional structures. It is important to mention that we also performed calculations using the original B3LYP functional without dispersion corrections and found that these geometry optimizations resulted in unstable and unbound nanotubes (a previous study by one of us also found that B3LYP yields purely repulsive interactions in fullerene-encapsulated nanostructures[29]). We also investigated other rotational orientations between adjacent monomers and found that our periodic geometries had similar structures to the most stable conformations in the molecular study by Flores *et al.*[58] Figures, Cartesian coordinates, and total energies for all of the optimized *syn* and *anti* nanotubes can be found in the EPAPS.[68] Geometries of C*n*T-*syn* nanotubes were calculated for *n* = 6-12, and C*n*T-*anti* nanotubes were calculated for *n* = 8, 10, and 12 (note that a complete *anti*-conformation cannot be obtained if a ring contains an odd number of monomers). For the C*n*T-*anti* nanotubes, each cyclic monomer is oriented with its thiophene rings parallel to the cylindrical tube axis. Table III compares the geometric and electronic properties of the C*n*T-*anti* monomers and nanotubes as a function of monomer size. In the optimized geometries, we find that monomers are repeated along the tube axis with very little axial rotation relative to each other, and the dihedral angle between thiophene rings is fairly constant at ~150° for all of the *anti*-conformations. Since there is no π-π stacking in the *anti* configuration, the cohesive energies for these nanotubes grow very weakly as a function of size (at a rate of 0.04 eV/monomer size).

In the C*n*T-*syn* nanotubes, the cyclic monomers are oriented with their thiophene rings perpendicular to the tube axis. Table IV summarizes the geometric and electronic properties of the C*n*T-*syn* monomers and nanotubes as a function of monomer size. The monomer to monomer repeat distances for the $n$ = 6-12 C*n*T-*syn* nanotubes are within 3.5-4.3 Å, which are in accordance to average π-π stacking distances of 3.2-3.8 Å in aromatic macrocycles.[71] These inter-monomer distances also allow significant electron delocalization between monomers, as demonstrated in a recent experimental-theoretical study[13,14] which showed that electron tunneling between adjacent monomers (as quantified by the Marcus transfer integral, *J*) is still very high even at inter-monomer distances of 3.4 Å. Furthermore, as the ring size of each monomer increases, the cohesive energy becomes more stabilized from 0.52 eV to 2.43 eV. These cohesive energies are significantly more stable than the corresponding C*n*T-*anti* geometries since they include both dispersion and π-π stacking interactions between adjacent monomers. As a result, the strong π-π interactions in the C*n*T-*syn* nanotubes provide an extra source of electronic stability in their self-assembly. It is also interesting to note that the *syn* monomers themselves (i.e. not in a periodic geometry) are more stable than the corresponding *anti* monomers by 0.29, 0.73, and 0.85 eV for the $n$ = 8, 10, and 12 monomers respectively (total energies for all of our monomers can be found in the EPAPS[68]). The additional stability of the *syn* monomers is due to reduced strain energies and is consistent with the theoretical study by Zade and Bendikov which investigated isolated cyclic monomers.[72] Furthermore, in contrast to the C$n$T-*anti* nanotubes, adjacent monomers in the *syn* configuration are offset from each other by a slight rotation around the tube axis (see Fig. 1a). This construction is favored due to a balance between electrostatic repulsion between nuclei, as well as favorable $\pi$-$\pi$ interactions that result from the delocalized electrons between macrocycles. Cyclic oligothiophenes consisting of 10-12 thiophene subunits are nearly flat due to nearly unstrained ring conformations, whereas smaller oligothiophenes have alternating dihedral angles away from planarity between adjacent thiophene rings (see Figs. E-4 – E-10 in the EPAPS[68]). Furthermore, due to the parity of the alternating bending structure, odd-numbered C7T- and C9T-*syn* nanotubes are forced to form irregular structures. Specifically, C9T-*syn* contains two adjacent thiophene rings bending in the same direction, and C7T-*syn* adopts a very strained conical structure. These irregular structures have a negative impact on $\pi$ orbital overlap and thus the overall stacking energetics.

**Electronic Band Structures.** To provide further insight into electronic properties, we plot the B3LYP-D band structure along the irreducible Brillouin zone (defined by the high-symmetry points $\Gamma$ and X in *k*-space) for the cyclic oligothiophene nanotubes in Figs. 2 and 3. The unit cell and the direction of the *k*-vector for a C9T-*syn* nanotube are shown in Fig. 4a. In all of the different conformations, we find that the electronic band structures yield a semiconducting behavior with a direct band gap at the $\Gamma$ symmetry point. However, the C$n$T-*anti* nanotubes have very large band gaps of about 3 eV which remain relatively constant across different ring sizes (Fig. 2). This result is expected since the width of an electronic band reflects orbital interactions along the nanotube, with wide bands denoting delocalization and narrow bands corresponding to localization/small orbital overlap. Since none of the C$n$T-*anti* nanotubes have favorable $\pi$-orbital overlap between monomers, all of the electronic bands are nearly dispersionless (non-interacting) and the band gap does not change dramatically with ring size.

Using our *ab initio* calculations for the C$n$T-*anti* nanotubes, we performed a nonlinear fit of the band gap ($E_g$) as a function of monomer size ($n_{rings}$) or diameter ($d$). We chose a flexible functional form given by $E_g = A/n_{rings} + B$, where $A$ and $B$ are independent free parameters subject to our nonlinear least-squares fit. Based on our B3LYP-D band gaps, we obtained a fitted expression given by

$$E_g\left(anti\right) = \frac{10.3 \text{ eV}}{n_{rings}} + 2.0 \text{ eV}$$
$$= \frac{1.3 \text{ eV}}{d \text{ (in nm)}} + 2.0 \text{ eV}. \qquad (6)$$

It is very interesting to note that the constant term in Eq. (6) corresponds to a band gap limit of a system where $n_{rings}$ (or $d$) is taken to infinity. In other words, the constant term in Eq. (6) yields the band gap of an *anti* nanotube having an infinite diameter (or an infinite number of rings). Although we determined this constant as a free parameter in our fit, it is noteworthy to point out that we recover the 2.0 eV band gap of periodic polythiophene (which is the limiting case of an *anti* nanotube with infinite diameter) calculated earlier in Table II.

In contrast to the *anti* nanotubes, we find that the electronic band structures for the C$n$T-*syn* tubes are qualitatively more similar to semiconducting carbon nanotubes, even though the monomers in the self-assembled columnar structures are *not* covalently bonded. The C$n$T-*syn* nanotubes have a direct band gap which decreases rapidly from 3.0 eV to 0.9 eV with increasing nanotube diameter, as shown in Fig. 3. In particular, we draw attention to the rapid decrease in band gap as the number of thiophene rings increases from 9 to 10 in the periodic C$n$T-*syn* nanotube. As mentioned in the previous section on cohesive energies, a structural-geometry transition in the nanotubes occurs when monomer subunits with $n \geq 10$ become flat due to unstrained ring conformations, whereas smaller nanotubes with $n \leq 9$ have monomers with alternating dihedral angles away from planarity between adjacent thiophene rings (see Figs. E-4 – E-10 in the EPAPS[68]). It is interesting to note that all of the isolated monomers up to $n = 12$ never achieve complete planarity in our geometry optimizations, resulting in a rather gradual variation in the HOMO-LUMO gap as a function of size, as shown in Table III. As a result, the significant decrease in band gap is unique to the one-dimensional C$n$T-*syn* nanotube geometry, leading to an enhanced electron delocalization which is not available in the isolated monomer case. Using the same functional form as Eq. (6), we also performed a nonlinear least-squares fit of the band gap as a function of monomer size for the C$n$T-*syn* nanotubes. From B3LYP-D calculations for the larger nanotubes with $n_{\text{rings}} \geq 10$ (the smaller C$n$T-*syn* tubes are excluded from the fit since they have very high ring strain), we obtained a fitted expression given by

$$E_g(syn) = \frac{8.5 \text{ eV}}{n_{\text{rings}}} + 0.2 \text{ eV}$$
$$= \frac{1.1 \text{ eV}}{d \text{ (in nm)}} + 0.2 \text{ eV}. \tag{7}$$

Again, the constant term in Eq. (7) corresponds to a band gap limit of a *syn* nanotube having an infinite diameter. In this case, the infinite-diameter limit corresponds to a sheet composed of polythiophene polymers $\pi$-stacked (and still strongly interacting) with each other (this is in contrast to the infinite-diameter limit of an *anti* nanotube, which yields a sheet composed of polymers not interacting with each other since there is no $\pi$-stacking in the *anti* configuration; cf. Fig. 1). Finally, it is important to point out that both Eq. (6) and Eq. (7) differ from the empirical expression for semiconducting carbon nanotubes[73,74] ($E_g = 0.84/d$ (nm), where $d$ is the diameter of a non-metallic nanotube) since each of our expressions have a constant energy term implying that cyclic oligothiophene nanotubes are always semiconducting and will not give a zero band gap regardless of size. This result is expected since conductivity in these noncovalently-bound nanotubes will be ultimately limited by hopping transport between $\pi$ orbitals of adjacent monomers, which we estimate in Eq. (7) to have a limiting band gap value of 0.2 eV.

## IV. CONCLUSION

In this study, we have investigated the band structure and size-scaling of electronic properties in a series of cyclic oligothiophenes which self-assemble to form stable nanotubes. Using a dispersion-corrected B3LYP-D hybrid functional which simultaneously incorporates noncovalent interactions and reduced self-interaction error, we find that the *anti* and *syn* oligothiophene nanotubes demonstrate very different electronic properties and stabilities. Cyclic oligothiophenes assembled in the *anti* configuration form weakly-bound, large band gap nanotube structures with dispersionless/non-interacting electronic bands. In contrast, cyclic oligothiophene nanotubes in the *syn* configuration form extremely stable, delocalized $\pi$-stacked structures with tunable band gaps as a function of size. Simple formulas for estimating the fundamental band gaps in both the *anti* and *syn* nanotubes are presented as a function of size and diameter [Eqs. (6) and (7)]. Most importantly, we find that the *syn* nanotubes have additional $\pi$-$\pi$ stacking energies and favorable geometric relaxation, resulting in very stable tubular aggregates of cyclic oligothiophenes held together purely via noncovalent interactions. This combination of stability and electron delocalization in the *syn* nanotubes is favorable towards one-dimensional electron transport and energy migration along the nanotube axis.

In conclusion, these results suggest that cyclic oligothiophene nanotubes can be spontaneously formed from facile self-assembly and can be used as nanoscale semiconducting materials with tunable electronic and geometric properties. Furthermore, the self-assembly of identical molecular building blocks into discrete, one-dimensional nanostructures is a powerful strategy for producing nanomaterials having a well-defined electronic structure (as opposed to carbon-nanotube production which still requires extensive sorting of different electronic types). Although we have only focused on the self-organization of well-defined cyclic oligothiophenes into nanotubes, chemical functionalization of these columnar structures via covalent cross-linking or noncovalent attachment of photochromic molecules[75-77] to the nanotube walls can provide a mechanism for further modification of the nanotube band structure.[78,79] Alternatively, incorporation of molecules such as $C_{60}$ within the spacious nanotube cross-section can cause charge-carrier formation, which would further enhance electron mobility in photovoltaic devices, nanosensors, and organic transistors.[80] We are currently investigating these optoelectronic properties within the framework of time-dependent DFT[81-85] and the two-particle Bethe-Salpeter equation[86-89] which are necessary for accurate descriptions of these optical processes.

## ACKNOWLEDGMENT

We acknowledge helpful discussions with Andrew L. Vance. This research was supported in part by the National Science Foundation through TeraGrid resources (Grant No. TG-CHE1000066N) provided by the National Center for Supercomputing Applications. Funding for this effort was provided by the Laboratory Directed Research and Devlopment (LDRD) program at Sandia National Laboratories, a multiprogram laboratory operated by Sandia Corporation, a Lockheed Martin Company, for the United States Department of Energy under contract DE-AC04-94AL85000.

.

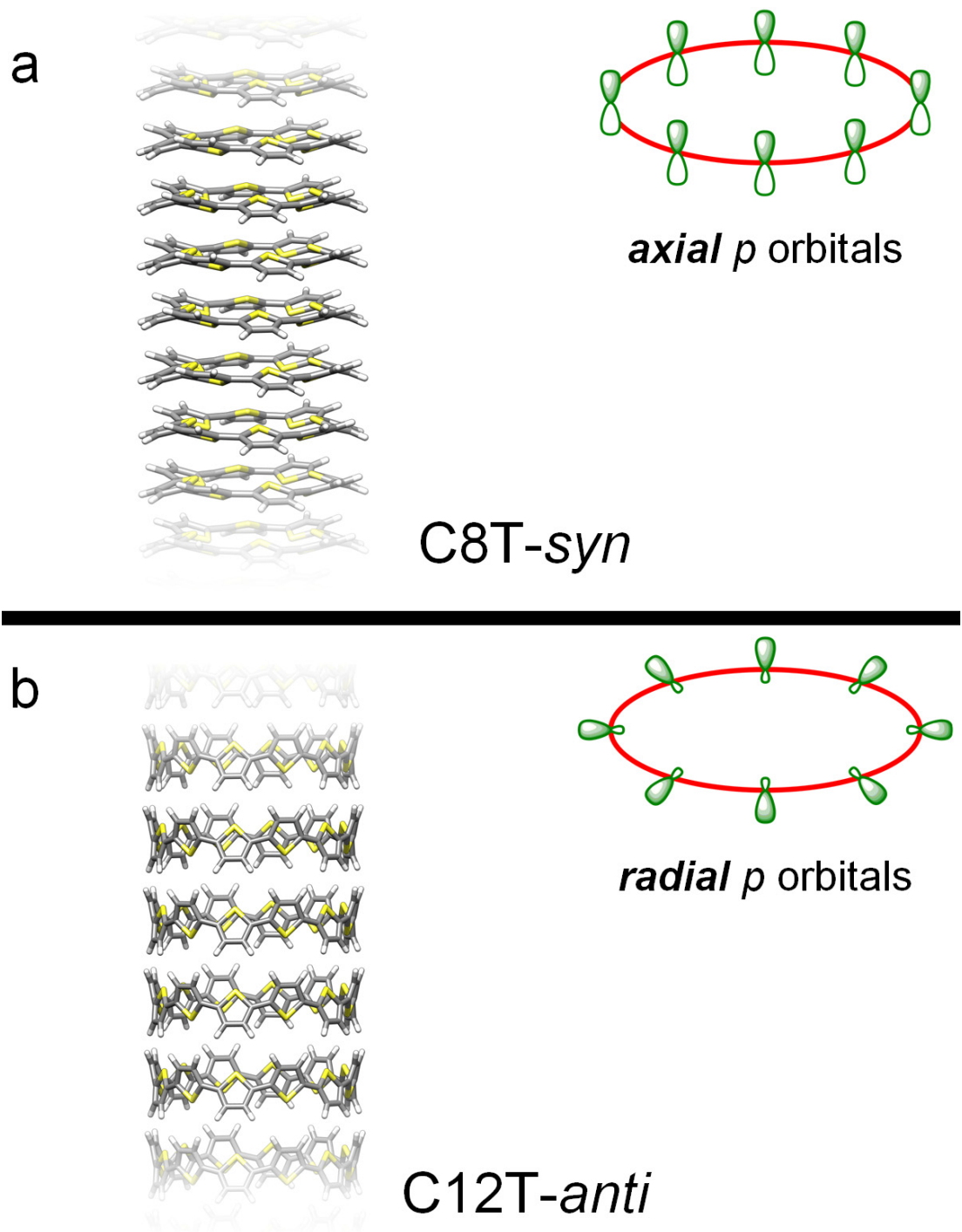

FIG. 1. (Color online) Cyclic oligothiophene nanotubes consisting of (a) $n = 8$ thiophene subunits in the *syn* configuration, and (b) $n = 12$ thiophene subunits in the *anti* configuration. In the *syn* configuration, the thiophene *p* orbitals are oriented along the axis of the nanotube, while the thiophene *p* orbitals point radially outward in the *anti* configuration. Also note that adjacent monomers in the *syn* configuration are offset from each other by a slight rotation around the tube axis.

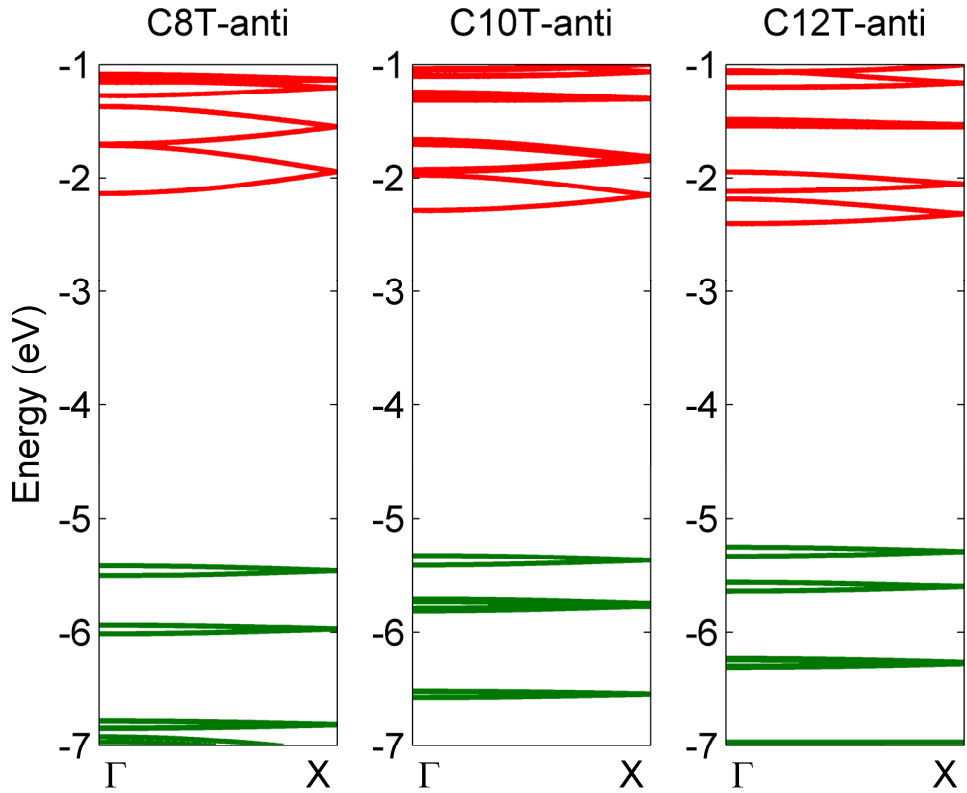

FIG. 2. (Color online) Electronic band structures (relative to vacuum at 0 eV) of the C$n$T-*anti* nanotubes for $n$ = 8, 10, and 12. All of the electronic bands are nearly dispersionless (non-interacting), and the band gap does not change dramatically with ring size.

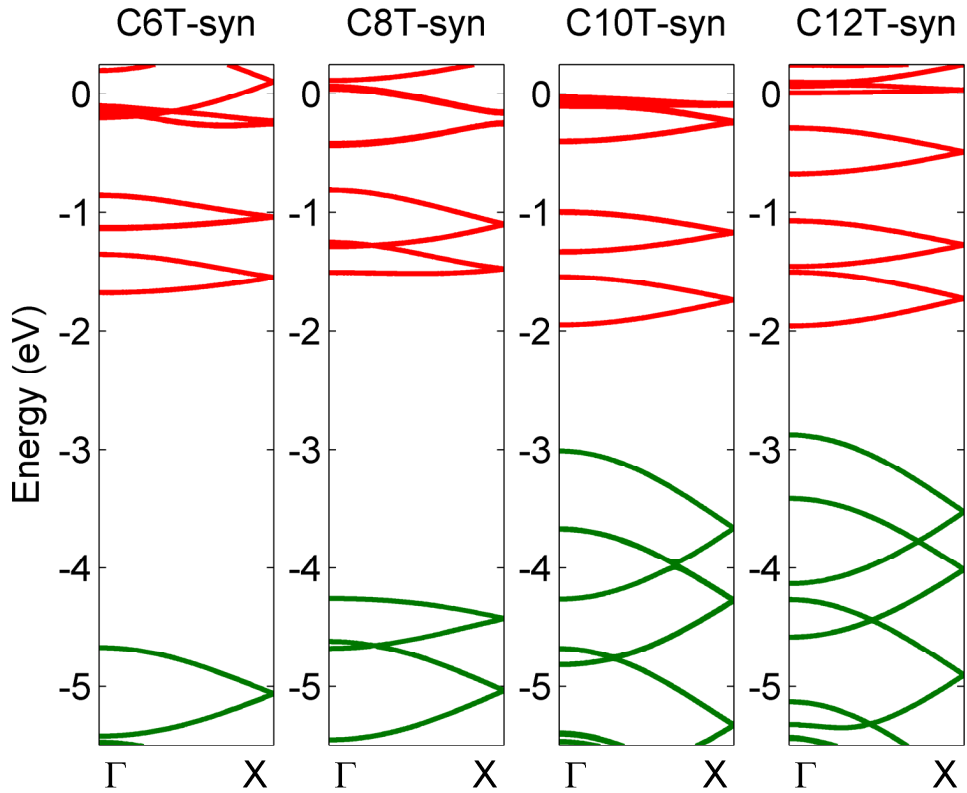

FIG. 3. (Color online) Electronic band structures (relative to vacuum at 0 eV) of the C$n$T-*syn* nanotubes for $n$ = 6, 8, 10, and 12. The direct band gap at the Γ-point decreases rapidly with increasing nanotube diameter.

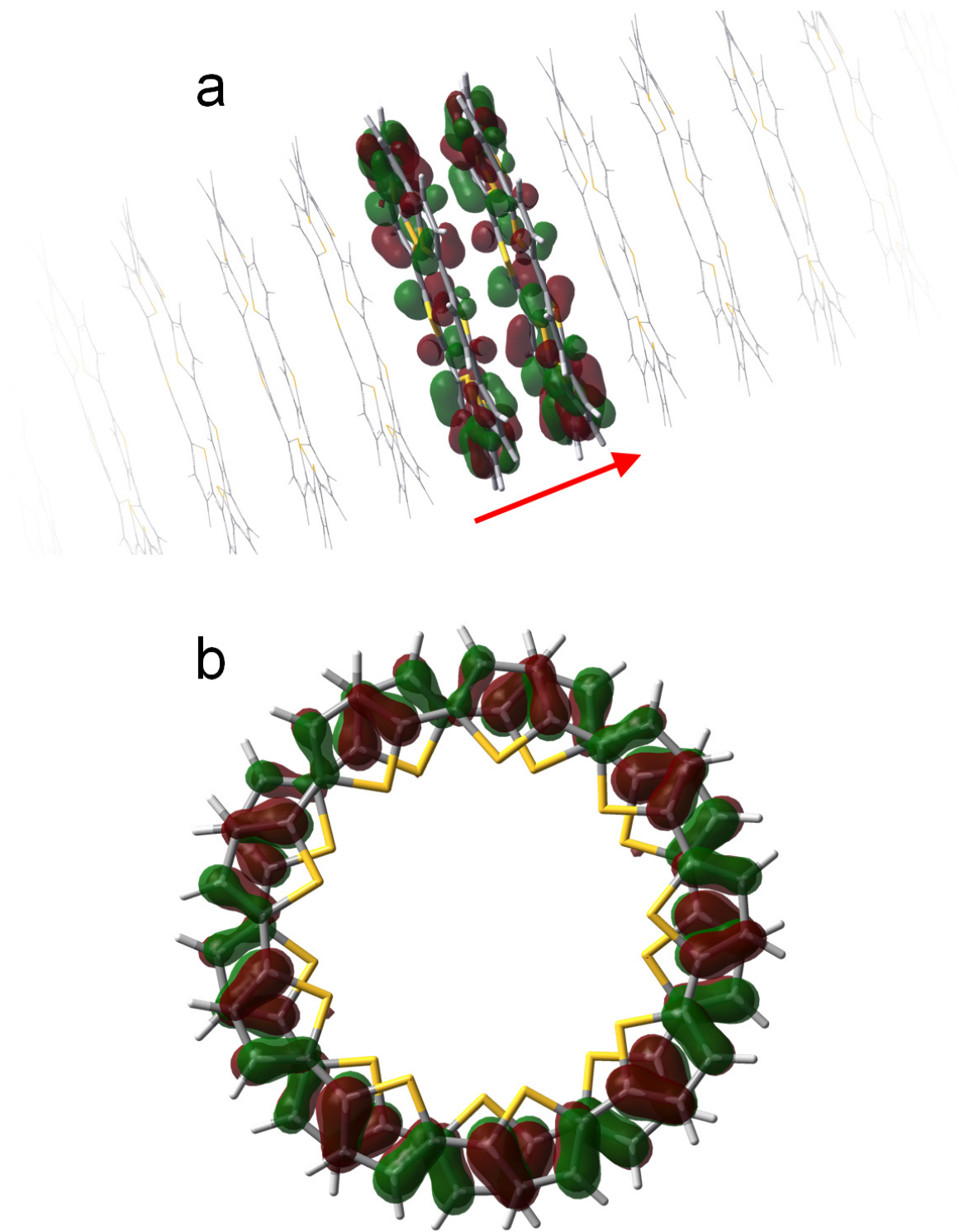

FIG. 4. (Color online) Highest occupied crystal orbitals (HOCO) for the C9T-*syn* nanotube as viewed (a) along the side and (b) along the axis of the nanotube. The arrow shown in (a) denotes the translation vector for the one-dimensional periodic unit cell used in the B3LYP-D calculations. Monomer repeat units in (b) have been omitted for clarity.

TABLE I. Cohesive energies of oligothiophene monomers in the herringbone packing structure. All B3LYP-D energies were calculated using the all-electron 6-311G(d,p) basis using B3LYP-D/6-31G(d,p)-optimized geometries.

|                                | Cohesive Energy (eV) | | |
|--------------------------------|--------|---------|---------|
| $n$, number of thiophene rings | LDA[a] | B3LYP-D | vdW-DF[a] |
| 2                              | 0.8    | 1.1     | 1.0     |
| 4                              | 1.3    | 2.0     | 1.8     |
| 6                              | 1.9    | 2.9     | 2.8     |

[a]Reference 57.

TABLE II. Electronic band gaps of periodic polythiophene. All band gaps were calculated at the B3LYP-D/6-311G(d,p) level of theory using B3LYP-D/6-31G(d,p)-optimized geometries.

|  | Band Gap (eV) |
|---|---|
| LDA | 1.05 |
| BLYP | 0.96 |
| B3LYP-D | 1.95 |
| Experimental[69] | 2.00 |

TABLE III. HOMO-LUMO gaps for isolated C$n$T-*anti* oligothiophene monomers and inter-monomer distances, nanotube diameters, cohesive energies, and electronic band gaps for periodic C$n$T-*anti* nanotubes. All calculations utilized the 6-311G(d,p) basis at the B3LYP-D level of theory.

| | C$n$T-*anti* monomer | C$n$T-*anti* periodic nanotube | | | |
|---|---|---|---|---|---|
| $n$, number of thiophene rings | HOMO-LUMO Gap (eV) | Inter-monomer Distance (Å) | Nanotube Diameter (Å) | Cohesive Energy (eV) | Band Gap (eV) |
| 8 | 3.37 | 5.7 | 10.1 | 0.54 | 3.28 |
| 10 | 3.05 | 5.8 | 12.6 | 0.61 | 3.04 |
| 12 | 2.83 | 5.9 | 15.1 | 0.70 | 2.85 |

TABLE IV. HOMO-LUMO gaps for isolated C*n*T-*syn* oligothiophene monomers and inter-monomer distances, nanotube diameters, cohesive energies, and electronic band gaps for periodic C*n*T-*syn* nanotubes. All calculations utilized the 6-311G(d,p) basis at the B3LYP-D level of theory.

|  | C*n*T-*syn* monomer | C*n*T-*syn* periodic nanotube | | | |
| --- | --- | --- | --- | --- | --- |
| *n*, number of thiophene rings | HOMO-LUMO Gap (eV) | Inter-monomer Distance (Å) | Nanotube Diameter (Å) | Cohesive Energy (eV) | Band Gap (eV) |
| 6 | 3.62 | 3.9 | 7.9 | 0.52 | 2.99 |
| 7 | 2.56 | 4.3 | 9.2 | 1.29 | 2.55 |
| 8 | 2.84 | 3.7 | 10.4 | 1.70 | 2.75 |
| 9 | 2.58 | 3.6 | 11.7 | 1.91 | 2.58 |
| 10 | 2.34 | 3.5 | 13.0 | 1.91 | 1.06 |
| 11 | 2.16 | 3.5 | 14.3 | 2.19 | 0.99 |
| 12 | 2.17 | 3.5 | 15.5 | 2.43 | 0.92 |